\documentclass[journal=jacsat,manuscript=article]{achemso}

\usepackage[version=3]{mhchem} 
\usepackage[T1]{fontenc}       
\usepackage{graphicx}
\usepackage{caption}
\author{Ronan Gourgues}
\affiliation[SQ]
{Single Quantum B.V., 2628 CH Delft, The Netherlands.}

\email{ronan@singlequantum.com}

\author{Iman Esmaeil Zadeh}
\affiliation[Optics]
{Optics Research Group, ImPhys Department, Faculty of Applied Sciences, Delft University of Technology, Lorentzweg 1, 2628 CJ Delft, The Netherlands.}
\alsoaffiliation[SQ]
{Single Quantum B.V., 2628 CH Delft, The Netherlands.}

\author{Ali W. Elshaari}
\affiliation[KTH]
{Department of Applied Physics, Royal Institute of Technology (KTH), SE-106 91 Stockholm, Sweden.}

\author{Gabriele Bulgarini}
\affiliation[SQ]
{Single Quantum B.V., 2628 CH Delft, The Netherlands.}

\author{Johannes W. N. Los}
\affiliation[SQ]
{Single Quantum B.V., 2628 CH Delft, The Netherlands.}

\author{Julien Zichi}
\affiliation[KTH]
{Department of Applied Physics, Royal Institute of Technology (KTH), SE-106 91 Stockholm, Sweden.}

\author{Dan Dalacu}
\affiliation[Ottawa]
{National Research Council of Canada, Ottawa, ON K1A 0R6, Canada.}

\author{Philip J.Poole}
\affiliation[Ottawa]
{National Research Council of Canada, Ottawa, ON K1A 0R6, Canada.}

\author{Sander N. Dorenbos}
\affiliation[SQ]
{Single Quantum B.V., 2628 CH Delft, The Netherlands.}

\author{Val Zwiller}
\affiliation[KTH]
{Department of Applied Physics, Royal Institute of Technology (KTH), SE-106 91 Stockholm, Sweden.}
\alsoaffiliation[SQ]
{Single Quantum B.V., 2628 CH Delft, The Netherlands.}

\title[An \textsf{achemso} demo]
{Controlled integration of selected detectors and emitters in photonic integrated circuits}

\begin{document}

	\begin{abstract}
Integration of superconducting nanowire single photon detectors and
quantum sources with photonic waveguides is crucial for realizing advanced quantum integrated circuits. However, scalability is hindered by stringent requirements on high performance detectors. Here we overcome the yield limitation by controlled coupling of photonic channels to pre-selected detectors based on measuring critical current, timing resolution, and detection efficiency.  As a proof of concept of our approach, we demonstrate a hybrid on-chip full-transceiver consisting of a deterministically integrated detector coupled to a selected nanowire quantum dot through a filtering circuit made of a silicon nitride waveguide and a ring resonator filter, delivering 100 dB suppression of the excitation laser. In addition, we perform extensive testing of the detectors before and after integration in the photonic circuit and show that the high performance of the superconducting nanowire detectors, including timing jitter down to 23 $\pm$ 3 ps, is maintained. Our approach is fully compatible with wafer level automated testing in a cleanroom environment.
\end{abstract}
	
\section{Introduction}

From visible to mid-infrared, superconducting nanowire single photon detectors (SNSPDs) have demonstrated excellent performances in terms of efficiency\cite{Marsili2013,doi:10.1063/1.5000001,Zhang2017}, temporal resolution\cite{2018arXiv180106574E,Korzh:2018zkh}, dark counts\cite{Schuck2013a} and detection rates\cite{doi:10.1063/1.5000001}. The fact that SNSPDs are compact and require only a single lithographic step for their realization allows for building large scale optical circuits. All these features make SNSPDs ideal candidates for photon detection in advanced quantum \cite{O'Brien2009} and neuromorphic integrated optical circuits \cite{Shen2017,PhysRevApplied.7.034013}. Waveguide-integrated SNSPDs are particularly interesting, because the evanescent coupling allows for close to unity detection efficiencies\cite{Kovalyuk:13,Pernice2012}, while still maintaining short detection length due to the strong coupling between the SNSPD and the guided optical mode. Several superconducting materials such as NbN and NbTiN have been used to fabricate detectors in a wide variety of photonic platforms: silicon-on-insulator\cite{Pernice2012}, silicon nitride-on-insulator\cite{Schuck2013}, diamond\cite{Rath2015}, and GaAs/AlGaAs \cite{Reithmaier2013}. The scalability of complex quantum photonic integrated circuits is limited by the fabrication yield of each component \cite{Metcalf2013}. High performance single-photon detectors based on  SNSPD technology are demanding due to challenges in sputtering high quality superconducting films and imperfections in the nanowire during lithography and etching steps. Constrictions along the nanowire affect the performance of the detector by limiting the device switching current well below its theoretical critical current \cite{doi:10.1063/1.2696926}. Due to these subtle non-determinisms in fabrication, without careful characterization, it is difficult to predict if all detectors meet the required performances.
In this work, we report on a deterministic approach for integrating high performance SNSPDs with photonic circuits. Additionally, we demonstrate an on-chip full transceiver consisting of a source, an optical link, and a detector on the same circuit. In the following sections we discuss the fabrication process and present the results.

\section{SNSPDs fabrication and characterization}

\begin{figure}
	\centering
	\includegraphics[scale=0.5]{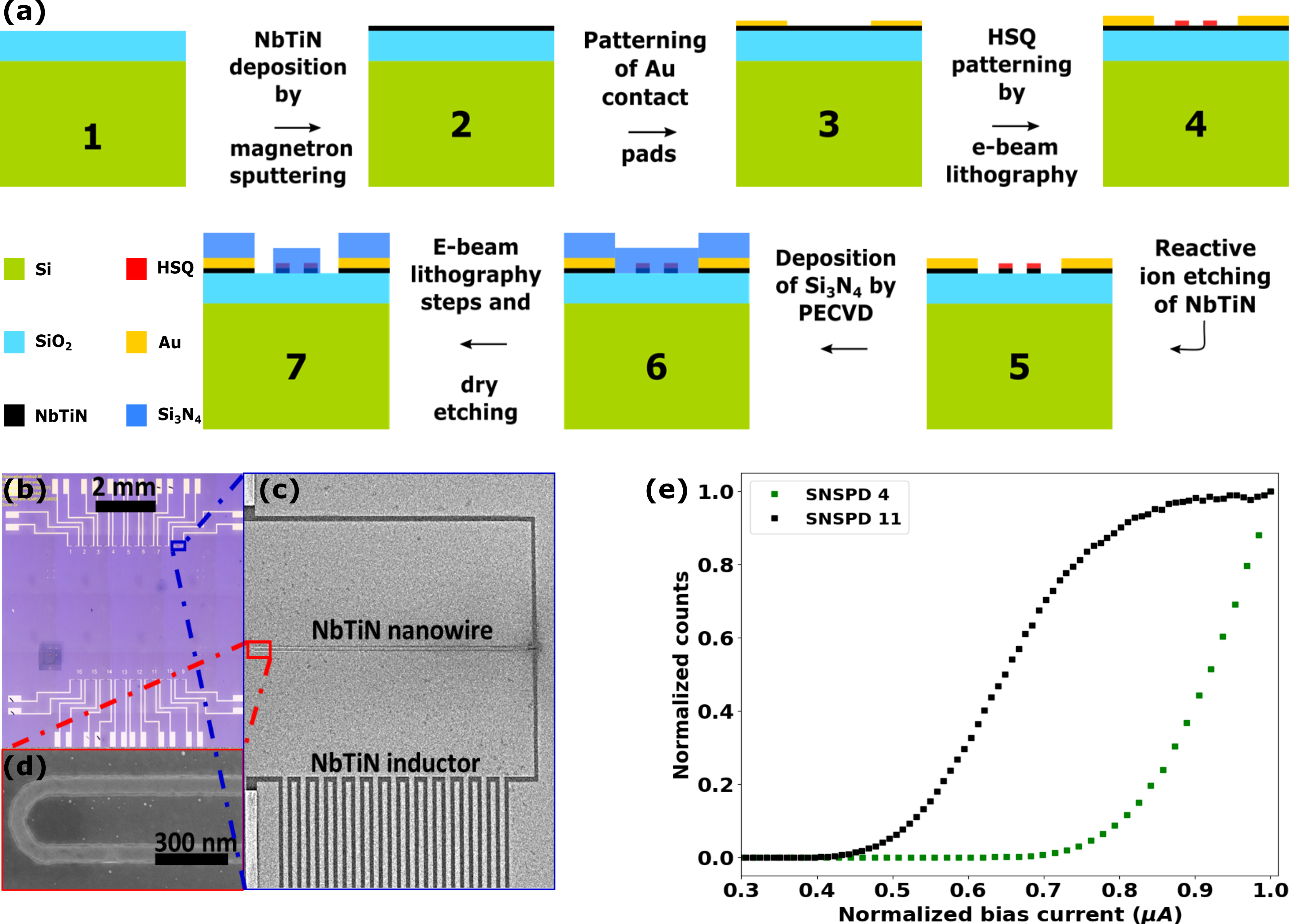}
	\caption{(a) Schematics of the fabrication process.
		(b) Optical microscope image of a chip with 16 SNSPDs.
		(c) SEM image of the SNSPD and the series inductor. 
		(d) SEM image of the U-shaped nanowire detector. 
		(e) Normalized detection efficiency at 881 nm for two detectors on the same chip VS bias current.}
	\label{fig:fig1}
\end{figure}

The process for fabricating the devices is shown in Figure\,~\ref{fig:fig1}a. We started with a silicon wafer and thermally oxidized it to form 3.6 $\mu$m SiO$_{2}$, which serves as the bottom cladding for the photonic waveguide [Figure\,~\ref{fig:fig1}(a1)]. Next, we sputtered  9.5 nm thick NbTiN using magnetron co-sputtering in an Ar and N\textsubscript{2} atmosphere [Figure\,~\ref{fig:fig1}(a2)]. The film thickness and composition were optimized to yield high critical current with saturated internal efficiency at $\sim$ 900 nm, close to the emission wavelength of selected nanowire quantum dots in the experiment at 4.2 K. Subsequently, Cr/Au contacts were formed using e-beam-lithography, evaporation and lift-off [Figure\,~\ref{fig:fig1}(a3)]. The nanowires were patterned on hydrogen silsesquioxane (HSQ) e-beam resist using 100 keV lithography system [Figure\,~\ref{fig:fig1}(a4)], then the pattern was transferred to the NbTiN layer by dry etching using SF$_{6}$ and O$_{2}$ chemistry [Figure\,~\ref{fig:fig1}(a5)]. The sample was mounted on a printed circuit board to allow electrical bias and read out. For the purpose of prototyping and due to the limited range of the translation stages, each chip houses 16 devices fabricated on a sample of size 1.8 $\times$ 1 cm$^{2}$ as shown in Figure\,~\ref{fig:fig1}b. Figure\,~\ref{fig:fig1}c presents a scanning electron microscope image of the detector. The active part of the detector, partly shown in Figure\,~\ref{fig:fig1}d, forms a  "U"-shape \cite{Pernice2012,Schuck2013,Rath2015,Khasminskaya2016} with  70 nm wide nanowires separated by a 200 nm gap. Additionally, to avoid latching due to the small kinetic inductance of the nanowire, we included a 2.5 mm long and 400 nm wide section serving as a series inductor \cite{0953-2048-22-5-055006}.
To characterize the detectors, the samples were directly immersed in liquid helium and illuminated from the top using attenuated CW laser at 881 nm. Detectors were biased using a tunable source and the detection events were counted by a high-speed counter.  

Figure\,~\ref{fig:fig1}e shows the normalized efficiency of two representative detectors on the same chip as a function of the bias current, the observation of a plateau in the count rate for fixed illumination indicates unity internal efficiency of the detector. We observe that the internal efficiency of detector $\boldsymbol{11}$ (black curve) saturates at approximately 80 $\%$ of the critical current, in sharp contrast with detector $\boldsymbol{4}$ (green curve). Detectors $\boldsymbol{11}$ and $\boldsymbol{4}$ have a critical current of 11.6 $\mu$A and 6.9 $\mu$A respectively. The low critical current and low internal efficiency of detector $\boldsymbol{4}$ demonstrate the necessity of selecting detectors when they are integrated with complex photonic circuits.

\section{Deterministic integration of the detectors}

\begin{figure}
	\centering
	\includegraphics[scale=0.5]{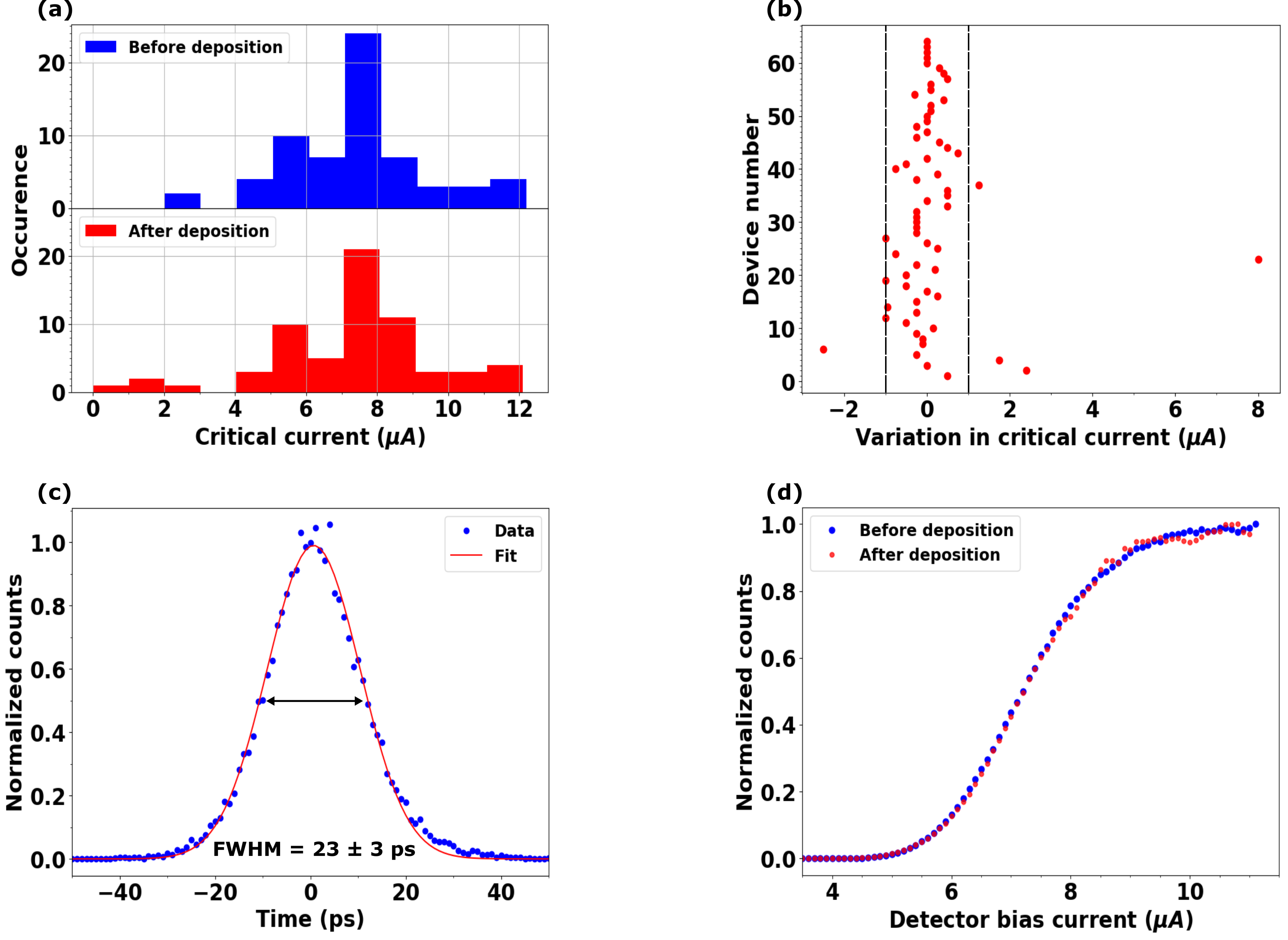}
	\caption{(a) Histograms of the critical current before (top) and after
		(bottom) deposition of Si$_{3}$N$_{4}$.  
		(b) Difference in critical current before and after deposition of Si$_{3}$N$_{4}$ for each detector, the two dashed vertical lines indicate current variation of -1 $\mu$A and 1 $\mu$A, respectively.       
		(c) Timing jitter measurement after deposition of Si$_{3}$N$_{4}$.
		(d) Detector normalized internal efficiency before and after deposition of Si$_{3}$N$_{4}$ as function of bias current.}
	\label{fig:fig2}
\end{figure}

Silicon nitride was selected for the waveguide core in our photonic circuit. It offers a wide optical transparency window from visible to mid-IR \cite{Rahim:17}, which makes it a good candidate for a range of photonic applications \cite{Elshaari2017,doi:10.1063/1.4962902,7463458}. Additionally, the relatively large refractive index-contrast with silicon oxide allows for single mode operation with high confinement of the optical mode \cite{6674990}. Furthermore, Si$_{3}$N$_{4}$ is deposited by plasma enhanced chemical vapor deposition (PECVD), the process has low thermal budget and CMOS compatible, making it suitable for large-scale CMOS-backend integration. To study the compatibility of the detectors with PECVD Si$_{3}$N$_{4}$, we initially characterized their critical current and internal optical detection efficiency before deposition. Next, we deposited a 200 nm thick layer of Si$_{3}$N$_{4}$ on top of the SNSPDs [Figure\,~\ref{fig:fig2}a6] and carefully re-measured each detector to study any deviation in performance. We used the same experimental set-up for all measurements.

A histogram of the critical current spread for 64 devices before and after PECVD deposition is shown in Figure\,~\ref{fig:fig2}a. The majority of devices have critical current in the range 7 $\mu$A - 8 $\mu$A. We calculated an average critical current of 7.52 $\mu$A before deposition and 7.42 $\mu$A after deposition. The standard deviation for the two distributions are 1.99 $\mu$A and 2.24 $\mu$A, respectively. The extracted statistics of the fabricated devices match those of commercially produced detectors for fiber-coupling, showing standard deviation of around one-fourth of the average critical current. We conclude from the analysis that photonic layer deposition has a negligible effect on the critical current of the detectors. We confirm this result by plotting the difference in critical current before and after deposition as shown in Figure\,~\ref{fig:fig2}b. The majority of detectors show a critical current variation of less than 1 $\mu$m. However, we notice larger changes in the critical current for a few devices, for instance devices 3 and 6 show a change in critical current by more than 2 $\mu$A. Out of 64 fabricated devices only one device shows open-circuit electrical response after deposition. Furthermore, we performed timing-jitter measurements for the highest performing detectors after etching the circuit. The experimental set-up used to perform the timing-jitter measurements is discussed in detail in \cite{doi:10.1063/1.5000001}. Figure\,~\ref{fig:fig2}c presents the temporal resolution of a selected device biased at 90 $\%$ of its critical current. The fitted data gives a full width half maximum (FWHM) timing-jitter = 23 $\pm$ 3 ps. A better time resolution can be achieved by using a cryogenic amplifier and by operating the SNSPD at a lower temperature which in turn increases the critical current and hence the signal to noise ratio \cite{2018arXiv180106574E}. Figure\,~\ref{fig:fig2}d shows the normalized efficiency versus bias current for the same detector. The critical current is 11.6 $\mu$A and 11.5 $\mu$A before and after deposition, respectively, the two curves overlap indicating a minimal influence of the deposition of silicon nitride on the performance of the detector. Based on timing-jitter, internal efficiency, and high critical current, we selected this detector for integration in a photonic circuit. The detector is coupled through a filtering circuit to a selected nanowire quantum dot to measure the quantum dot lifetime. The waveguides were patterned using e-beam lithography, the pattern was then transferred to Si$_{3}$N$_{4}$ by dry etching in a CHF$_{3}$/Ar chemistry [Figure\,~\ref{fig:fig2}a7].
To conclude this section, all extensive measurements reveal that testing the SNSPDs performance is needed to select the detectors before etching the waveguide layer, and the integrated circuit should be designed according to the selected SNSPDs.

\section{Quantum dot nanowire integration with SNSPDs}

\begin{figure}
	\centering
	\includegraphics[scale=0.5]{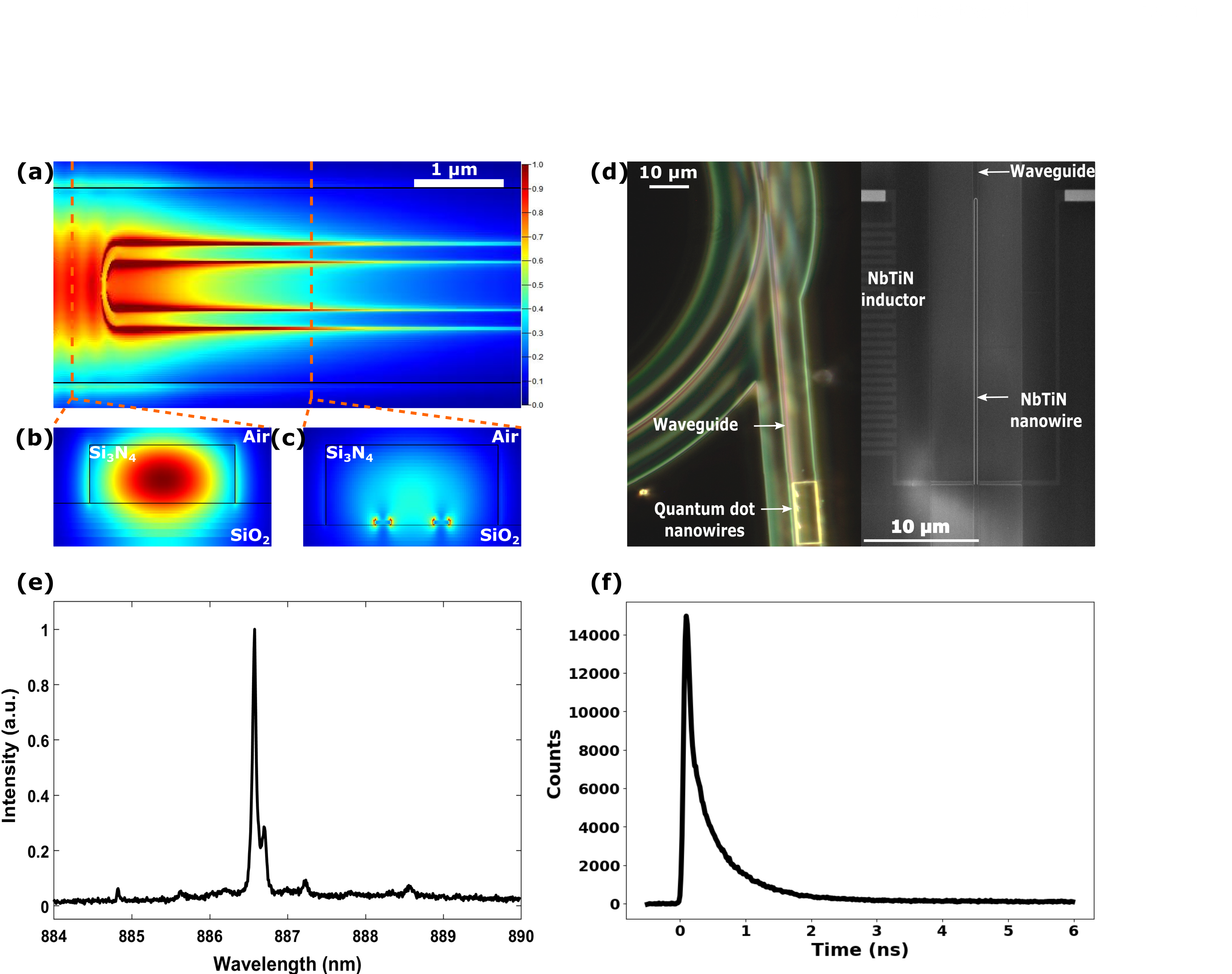}
	\caption{ (a) 3D FDTD simulation of near field intensity distribution (normalized) of the fundamental quasi-TE mode along 5 $\mu$m of NbTiN superconducting nanowire. The light is coupled to the waveguide from the left on the picture. (b) Simulated cross section of the electric field (normalized) in the Si$_{3}$N$_{4}$ waveguide before reaching the nanowire detector. (c) Simulated cross section of the electric field (normalized) in the Si$_{3}$N$_{4}$ waveguide with NbTiN nanowires after 2.5 $\mu$m of propagation.
		(d) Left : An optical picture of a part of the photonic circuit which includes quantum dots nanowires, a waveguide and a ring resonator. Right:  SEM image of the fabricated waveguide on top of the SNSPD.  
		(e) Spectrum of a selected quantum dot nanowire. 
		(f) Lifetime measurement of the quantum dot nanowire performed on chip.}
	\label{fig:fig3}
\end{figure}

Recent hybrid integration techniques \cite{Zadeh2016,Elshaari2017,Kim2017,Kim2018,Davanco2017} allow for combining high quality single photon emitters with silicon based photonic circuits. Waveguide coupled detectors are central elements in quantum photonic circuits to measure and analyse quantum states on chip. Efficient detection of the qubits is one of the main requirements for realizing the theoretical proposal for linear optical quantum computing\cite{Knill2001}, other requirements include ancillary states, post-selection, and feed-forward of the detection measurement. A key feature for waveguide integrated detectors is the high detection efficiency due to the strong evanescent coupling to the detector. To verify this, we perform three-dimensional finite-difference time-domain (3D FDTD) simulations. Figure\,~\ref{fig:fig3}a shows the optical mode absorption for a waveguide-coupled SNSPD. From the simulations we calculate an absorption coefficient of 3.1 dB/$\mu$m at the wavelength of 890 nm. The simulated waveguide matches the design of the fabricated chip with a height of 200 nm and  width of 800 nm. The optical mode propagating in the waveguide as shown in Figure\,~\ref{fig:fig3}b. The dimensions chosen for the waveguide provide a good confinement of the fundamental TE mode. Additionally, the selected air cladding delocalizes the optical mode into the substrate where the SNSPD is located, thus enhancing the SNSPD absorption by 10 $\%$ compared to the oxide cladding case. Figure\,~\ref{fig:fig3}c shows the eigen mode at the SNSPD region where the electric field intensity is maximum near the NbTiN wires. As a result, 91.4 $\%$ of the light is absorbed after 25 $\mu$m of propagation.

Based on the simulation results, we fabricated a full-quantum transceiver on-chip. The circuit consists of a ring resonator filter (Figure\,~\ref{fig:fig3}d left), the drop port of the filter is terminated by a waveguide coupled superconducting detector (Figure\,~\ref{fig:fig3}d right), while the input port is coupled to a nanowire quantum dot as shown in Figure\,~\ref{fig:fig3}d left. The quantum source was deterministically transferred using a nano-manipulation technique \cite{Zadeh2016,Elshaari2017} from the growth chip to the photonic circuit chip. The emission spectrum of the waveguide-coupled nanowire quantum dot is shown in Figure\,~\ref{fig:fig3}e. We designed the ring resonator so that the resonances have critical coupling around the emission wavelength of the quantum dot transition at 886.5 nm. The nanowire quantum dot was excited from the top using a femtosecond 515 nm pulsed laser with a repetition rate of 20 MHz. The wavelength of the laser was chosen to be within the absorption window of silicon nitride, thus the waveguide core acts as a natural high-pass filter to eliminate the pump photons which can blind the SNSPDs from detecting the QD signal. Additionally, the ring resonator is highly under-coupled for the pump laser photons, which provides an additional stage of filtering of the high intensity pump. The total suppression of the laser is estimated to be 100 dB. After excitation of the QD, the emitted photons are coupled to the silicon nitride waveguide, they are filtered by the ring resonator, and finally detected by the superconducting detector. Figure\,~\ref{fig:fig3}f shows a time-resolved start-stop correlation measurement with the laser signal, the SNSPD provides high time-resolution of  23 ps. We extracted the QD signal decay time of 0.62 $\pm$ 0.02 ns, in agreement with previous measurements performed on similar quantum dot nanowires configuration \cite{Zadeh2016,doi:10.1063/1.4948762}.

The presented measurement sets a standard for the level of determinism needed to realize large scale quantum photonic circuits, where detectors, as well as sources, are selected in a controlled process based on their individual characteristics.

\section{Conclusion}
In summary, we have shown a deterministic method to integrate high performance SNSPDs with photonic circuits. We realized on-chip full-transceiver, completely deterministic from source to detector and validated it by measuring the lifetime of a selected quantum dot. The integration process demonstrated in this article is CMOS compatible, detectors can be mass-produced and their characterization can be fully automated. Afterwards, only the best detectors are integrated with the photonic circuit. We believe that our method provides the needed accuracy and performance to realize future large scale quantum and neuromorphic photonic circuits.

\section{Funding}
A.W.E acknowledges support from Vetenskapsr\aa det Starting Grant (Ref: 2016-03905) and Marie-Sklodowska Curie Individual Fellowship under REA grant agreement No. 749971 (HyQuIP). I.E.Z acknowledges support from NWO LIFT-HTSM for Physics 2016-2017 under the  project number 680-91-202.
R.G  acknowledges  support  by  the  European  Commission  via  the  Marie-Sklodowska Curie action Phonsi (H2020-MSCA-ITN-642656).

\begin{acknowledgement}
We would like to thank Single Quantum for technical supports and Dr. Silvania Pereira for her contribution to this publication. 
\end{acknowledgement}

\bibliography{Ref}

\end{document}